\documentclass[trackchanges,twocolumn,twocolappendix]{aastex701}

\usepackage{amsmath}
\usepackage{graphicx,tabularx}	
 
\shorttitle{Mutual Information for Information Assessment}
\shortauthors{Sui et al.}
\submitjournal{ApJL}

\begin{document}
\title{How to evaluate the sufficiency and complementarity of summary statistics for cosmic fields: an information-theoretic perspective}

\correspondingauthor{Ce Sui, Yi Mao}

\author[0009-0004-5985-5123]{Ce Sui}
\affiliation{Department of Astronomy, Tsinghua University, Beijing 100084, China}
\email[show]{suic20@mails.tsinghua.edu.cn}  

\author[0000-0002-1301-3893]{Yi Mao} 
\affiliation{Department of Astronomy, Tsinghua University, Beijing 100084, China}
\email[show]{ymao@tsinghua.edu.cn}

\author[0000-0002-8328-1447]{Xiaosheng Zhao}
\affiliation{Department of Physics and Astronomy, Johns Hopkins University, Baltimore, MD 21218, USA}
\email{xzhao113@jh.edu}

\author[0009-0004-6271-4321]{Tao Jing}
\affiliation{Department of Astronomy, Tsinghua University, Beijing 100084, China}
\email{jingt20@mails.tsinghua.edu.cn}

\author[0000-0002-5854-8269]{Benjamin D. Wandelt}
\affiliation{Department of Physics and Astronomy, Johns Hopkins University, Baltimore, MD 21218, USA}
\affiliation{Department of Applied Mathematics and Statistics, Johns Hopkins University, Baltimore, MD 21218, USA}
\email{wandelt@jhu.edu}

\begin{abstract}
The advent of increasingly advanced surveys and cosmic tracers has motivated the development of new inference techniques and novel approaches to extracting information from cosmic fields. A central challenge in this endeavor is to quantify the information content carried by these summary statistics in cosmic fields. In particular, how should we assess which statistics are more informative than others and assess the exact degree of complementarity of the information from each statistic? Here, we introduce mutual information (MI) that provides, from an information-theoretic perspective, a natural framework for assessing the sufficiency and complementarity of summary statistics in cosmological data. We demonstrate how MI can be applied to typical inference tasks to make information-theoretic evaluations, using two representative examples: the cosmic microwave background map, from which the power spectrum extracts almost all information as is expected for a Gaussian random field, and the 21~cm brightness temperature map, from which the scattering transform extracts the most non-Gaussian information but is complementary to power spectrum and bispectrum. Our results suggest that MI offers a robust theoretical foundation for evaluating and improving summaries, thereby enabling a deeper understanding of cosmic fields from an information-theoretic perspective.
\end{abstract}

\keywords{\uat{Cosmology}{343} --- \uat{Bayesian statistics}{1900} --- \uat{Astrostatistics}{1882}}

\section{Introduction} 

Tracing the evolution of our Universe through different cosmological observations has been a central goal of modern cosmology. Among these endeavors, the cosmic microwave background (CMB) and the large-scale structure (LSS) stand out as the most precise observational probes. On large scales, both fields are well described as Gaussian random fields (GRFs), which means that their information content is fully captured by the two-point statistics, i.e. the power spectrum (PS) in Fourier space, which is therefore the focus of standard cosmological analysis (e.g.\citealt{Plank18,SDSS_cosmology,DESI_cosmology}). 

However, the anisotropies in the CMB and the nonlinear features of LSS, as well as new observational windows such as small-scale galaxy clustering, weak lensing, and 21~cm intensity mapping \citep[e.g.,][]{DESIY3,Euclid,SKA_cosmo}, contain additional information beyond the PS. To extract the non-Gaussian information from these datasets, significant efforts have been devoted to construct novel, informative summary statistics. Information extraction from these summaries is facilitated with the emergence of simulation-based inference \citep{ltu-ili,simbig,pydelfi}, because these statistics can be forward-modeled by advances in theories and increasingly realistic high-resolution simulations \citep{quijote,Flamingo,abacus,MTNG,IllustrisTNG,BaryonForge}. 
Examples of summary statistics include the use of cosmic voids as complementary probes to galaxy power spectra \citep{voids_Rosa,BOSS_voids,BOSS_voids_2}, bispectrum (BS) analyses across multiple cosmic fields \citep{Boss_bispectrum,WL_bispectrum,Watkinson_bs_eor_infer,Shimabukuro_bs_eor_infer}, and the application of Minkowski functionals to weak lensing and 21~cm maps \citep{MFs,MFs2,kangning_MF,WL_minks}. More recently, techniques such as the wavelet scattering transform (ST;  \citealt{cheng2020new}) have been employed to extract multiscale non-Gaussian information from various fields \citep{cheng2020new,xs-st-delfi,simbig_st}. In addition to such physically motivated summaries, machine learning–based approaches have been developed to construct ``optimal'' statistics, typically by maximizing an information criterion or improving inference performance \citep{hybrid,IMNN,xs-cnn-delfi}. 
Beyond the inference from summary statistics, field-level inference \citep{benchmark_fli,galaxy_fli,wl_fli} is also developed along the line to fully exploit non-Gaussian information in these datasets. 

In this era of thriving information, cosmologists seek to extract the {\it optimal} (sometimes aka {\it maximal} or {\it full}) information from different statistics, as well as providing complementary information by combining them. A key question arises: how should we assess quantitatively which statistics are more informative than others and assess the exact degree of complementarity of the information from each statistic? 

Assessing or learning summaries ultimately relies on quantifying their information content with respect to the underlying model parameters. In practice, this has often been approached through Fisher analysis \citep{fisher_for_21cm, data_compress_mi_cosmo}, posterior comparisons \citep{beyond2pt,DESI_st_valid} or regression performance \citep{21cm_CNN,21cmRNN}.  Among these, Fisher information (FI; \citealt{Tegmark_fisher}) is a widely used tool from estimation theory. However, its application in practice often relies on Gaussian likelihood assumptions and finite-difference score estimation, highlighting the need for new methods that enable more accurate estimation \citep{fsm,jaxcosmo}. In addition, FI only reflects local sensitivity around fiducial parameters. Posterior comparisons directly compare the credible region in parameter space for an inference around a fiducial test set, so they are also limited to local sensitivity. Regression-based metrics, while providing a global view, estimate the performance of point estimates, not the posteriors.  

In this paper, we propose the use of mutual information (MI), a fundamental concept in information theory and modern statistical inference \citep{info_book,info_stats_lecture}, to assess the sufficiency and complementarity of summary statistics in cosmological data. 
MI is defined to quantify the reduction in uncertainty of one random variable given knowledge of another, making it a natural metric for evaluating information content. In machine learning, MI has served as the theoretical basis for numerous representation-learning objectives \citep{infomax,infonce,representation_learning_review,multi_view_iob}. In cosmology, it has also been employed to design new summaries of cosmic fields \citep{vmim,data_compress_mi_cosmo,hybrid}. In this paper, however, we take a different perspective --- here, MI is used not as a loss function, but as a diagnostic tool to assess and decompose the information content of cosmological fields and their summaries. The mathematical properties of MI enable a principled evaluation of summary statistics, providing insights into their effectiveness, potential improvements, and the extent to which they capture the available information.

Note that MI, FI and regression errors are related through several interesting information-theoretic inequalities (see a brief review of these relations in Appendix~\ref{app:FI_MI_ineq}). However, a key distinction between them is that while FI depends on the specific fiducial parameter, MI does {\it not}, because it provides a global measure of information. We emphasize that both MI and FI are useful, complementary statistical tools --- MI as a metric for information assessment, and FI as a tool for direct forecast of parameter sensitivities. 

The remainder of this paper is organized as follows. Section~\ref{sec:method} introduces the core concepts and methodology. 
We apply MI to two representative cases in Section~\ref{sec:ex_set}, including CMB-like GRFs and 21~cm non-Gaussian maps. We make concluding remarks in Section~\ref{sec:conclusion}. Some technical details are left to Appendix~\ref{app:proof} (on MI and Bayesian sufficiency), Appendix~\ref{app:FI_MI_ineq} (on the relationships between MI, FI, and estimation mean squared error), and Appendix~\ref{app:summaries} (on the definition of summary statistics). 
A part of preliminary results were previously summarized by us in a conference proceeding paper \citep{mi_icml}. 

\section{Methodology}
\label{sec:method}

MI, a central concept in information theory and statistical inference, quantifies how much information one random variable contains about another. The MI $I(x; y)$ about two random variables $x$ and $y$ is defined as \citep{info_book}
\begin{equation}
I(x; y) = D_{\mathrm{KL}}\!\left(p(x,y) \,\big\|\, p(x)p(y)\right) 
\equiv \mathbb{E}_{p(x,y)}\!\left[\log \frac{p(x,y)}{p(x)p(y)}\right],
\label{eq:MI}
\end{equation}
where $p(x,y)$, $p(x)$, $p(y)$ denote their joint and marginal distributions, respectively. Here, the Kullback–Leibler (KL) divergence $D_{\mathrm{KL}}(P\,\big\|\,Q)$ is defined as the expectation value of $\log(P(x,y)/Q(x,y))$ over the joint distribution $P(x,y)$. Note that MI is symmetric, i.e.\ $I(x; y) = I(y; x)$. 

Intuitively, MI measures the discrepancy between the joint distribution and the product of the marginals using the KL divergence. In other words, MI capture how much uncertainty in a variable is reduced after the other is observed. 

\subsection{MI as a Metric of Information Sufficiency}

Statistical inference is all about estimating the parameters $\theta$ of a physical model given the observed data $x$. Within the Bayesian framework, this corresponds to inferring the posterior distribution $p(\theta|x)$. However, raw data $x$ are often high-dimensional and may contain a large amount of irrelevant or redundant information. To mitigate this, it is common to employ summary statistics, i.e., lower-dimensional representations $s=s(x)$ designed to retain the informative features of the data with respect to the parameters of interest. 

If summary statistics $s$ are sufficient, they must preserve all the information in $x$ that is relevant to $\theta$. The classical notion of sufficient statistics is defined by the condition, 
\begin{equation}
p(\theta|x) = p(\theta|s(x))\,,
\label{eq:suf_sta}
\end{equation}
which means that sufficient statistics $s(x)$ preserve all information about $\theta$ contained in the raw observed data $x$. 
In practice, while exact sufficiency is often unattainable, statistics that closely approximate sufficiency generally yield more accurate and reliable inference.

MI provides a natural characterization of sufficiency, because the sufficiency condition (Eq.~\ref{eq:suf_sta}) is equivalent (see this detail in Appendix~\ref{app:proof}) to 
\begin{equation}
I(\theta; x) = I(\theta; s(x))\,.
\label{eq:suf_sta_mi}
\end{equation}
Since the complete information from raw data $x$ is $I(\theta; x)$, the informativeness of summary statistics $s(x)$ can be evaluated using their MI $I(\theta; s(x))$ --- a higher value of MI means closer to satisfaction of the sufficiency condition and therefore the summary statistics extract more information from the data. This highlights MI as a powerful metric of information sufficiency, offering, from a solid and theoretical foundation, a practical criterion for guiding the design and selection of summary statistics in statistical inference.

\begin{figure}
    \centering
    \includegraphics[width=1\linewidth]{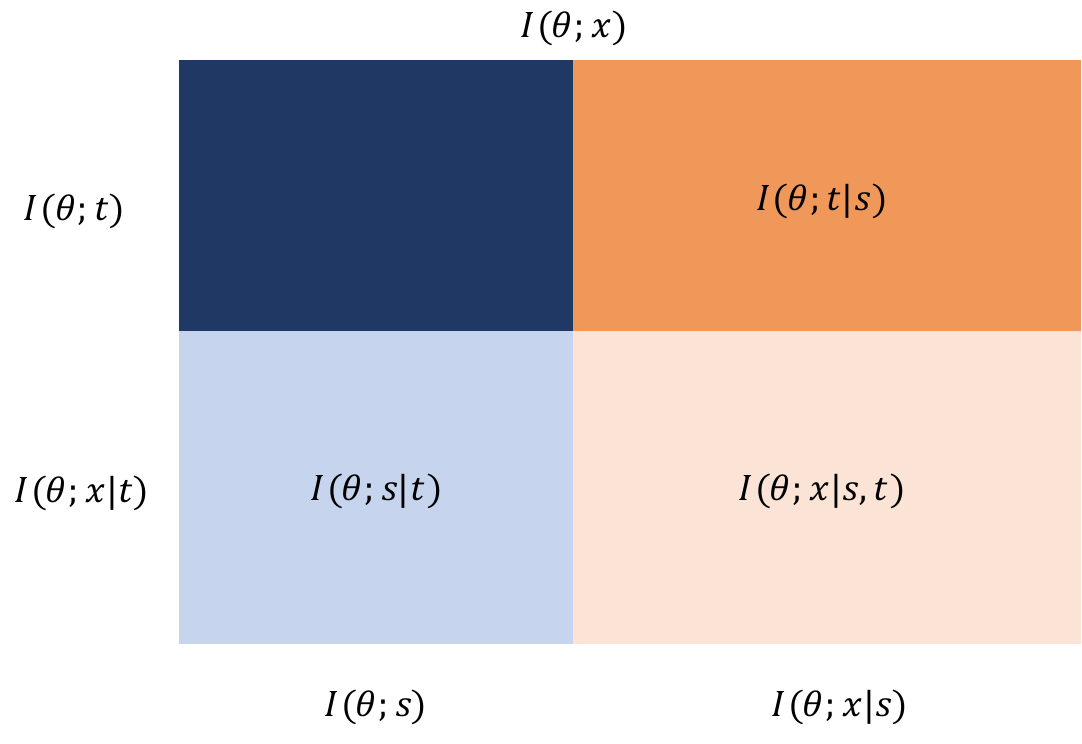}
    \caption{Decomposition of information content in a raw data $x$. $I(\theta;x)$ denotes the full field-level information. The left column (blue) $I(\theta;s)$ represents the information extracted by the summary $s$, and the right column (orange) $I(\theta ;x \mid s)$ captures the residual information beyond $s$. Similarly, the upper row (dark) $I(\theta ; t)$ shows the information extracted by another summary $t$, and the lower row (light) $I(\theta ;x \mid t)$ captures the residual information beyond $t$. The upper right (dark orange) panel $I(\theta ; t \mid s)$ quantifies the complementary information that $t$ contributes beyond $s$, and the lower left (light blue) panel $I(\theta ; s \mid t)$ quantifies the complementary information that $s$ contributes beyond $t$. The upper left (dark blue) panel quantifies the common information shared by both $s$ and $t$, and the lower right (light orange) panel $I(\theta ; x \mid s, t)$ quantifies the residual information in the raw data $x$ beyond both $s$ and $t$.}
    \label{fig:mi_diag}
\end{figure}

\subsection{Conditional MI as a Metric of Complementarity}

A particularly useful feature of MI is that it naturally allows us to decompose the information content of a field into contributions from different summaries. This provides insight into how summaries can be combined to maximize the captured information. This decomposition typically involves conditional MI, defined as
\begin{equation}
I(x; y \mid z) 
=\mathbb{E}_{p(x,y,z)}\!\left[\log \frac{p(x \mid y, z)}{p(x \mid z)}\right]\,,
\end{equation} 
where $x$, $y$ and $z$ are three random variables. 

The total information in the raw data $x$ can be decomposed using the chain rule of MI, 
\begin{equation}
I(\theta ; x ) 
= I( \theta ; s) + I(\theta ; x \mid s)\,,
\end{equation}
where the first term in the rhs represents the information captured by the summary statistics $s$, and the second term captures the residual information in the raw data $x$ beyond $s$.  

If two summary statistics $s$ and $t$ are considered, this decomposition can be further refined as
\begin{equation}
I(\theta ; x ) 
= I(\theta ; s) + I(\theta ; t \mid s) + I(\theta ; x \mid s, t)\,.
\label{eq:secdecomp}
\end{equation}
We illustrate this decomposition schematically in Figure~\ref{fig:mi_diag}. Here, $I(\theta ; t \mid s)$, corresponds to the complementary information that $t$ contributes beyond $s$. Conversely, $I(\theta ; s \mid t)$ corresponds to the complementary information that $s$ contributes beyond $t$. 

These two conditional MI can be used as the metric for quantifying the degree of information complementarity between two summary statistics, i.e.\ providing a direct measure of the additional information contributed by one summary beyond another base summary. 
Rather than replacing an existing summary, it may be more effective to identify complementary summaries that capture additional information --- an approach explored in \cite{hybrid}. 




\subsection{Numerical Estimation of MI}

Estimating MI in practice is challenging because it requires approximating high-dimensional probability densities. Classical approaches, e.g.\ the $k$-nearest neighbour estimator \citep{KSG}, approximate densities locally but do not scale well with dimensionality. The advent of neural networks has enabled flexible models capable of representing complex distributions or their ratios, giving rise to a range of MI estimators based on density modeling or density-ratio estimation \citep{MINE, VBofMI, SMILE}. Many of these can be formulated as learning objectives that find the extrema of MI \citep{infomax,infonce,multi_view_iob,MI_upper}.

In cosmological inference problems, the joint distribution of physical parameters and data summaries, $p(\theta,s)$, is often intractable. Consequently, direct computation of MI is infeasible. Instead, one can employ flexible generative models to approximate the true distributions and derive variational bounds on MI. 
If we introduce a variational distribution $q(\theta|s)$ to approximate the intractable conditional $p(\theta|s)$, then it is straightforward to show that 
\begin{equation}
I(\theta ;s ) = \mathbb{E}_{p(\theta, s)}\left[\log \frac{q(\theta | s)}{p(\theta)}\right] 
+ \mathbb{E}_{p(s)}\!\bigg[D_{\mathrm{KL}}\big(p(\theta | s)\,\|\,q(\theta | s)\big)\bigg]\,.
\label{eq:ba-bound}
\end{equation}

If $q(\theta|s)$ is sufficiently accurate such that $D_{\mathrm{KL}}(p(\theta | s)\,\|\,q(\theta | s)) \approx 0$, then MI can be approximated by
\begin{equation}
I(\theta ;s ) \approx \mathbb{E}_{p(\theta, s)}[\log q(\theta | s)] + h(\theta)\,.
\label{eqMIapprox}
\end{equation}
The first term in the rhs of Eq.~(\ref{eqMIapprox}) can be estimated by training a flexible generative model on large sets of parameter–statistic pairs to yield a variational distribution $q(\theta|s)$ that closely approximates the true posterior. In this paper, we employ a masked autoregressive flow (MAF; \citealt{MAF}) as a variational distribution. 
The second term in the rhs of Eq.~(\ref{eqMIapprox})  $h(\theta)=-\mathbb{E}_{p(\theta)}[\log p(\theta)]$ is the differential entropy of $\theta$. This corresponds to the Barber–Agakov lower bound \citep{ba-bound,VBofMI}. In cosmological applications, $h(\theta)$ is typically tractable since the prior on $\theta$ is known. 

The conditional MI can be estimated using chain rule,
\begin{equation}
I(\theta ; t \mid s) 
= I(\theta ; s,t) - I(\theta ; s)\,.
\label{eqconMI}
\end{equation}
The two terms on the rhs of Eq.~(\ref{eqconMI}) can be estimated by training two generative models, respectively. 

\section{Applications}
\label{sec:ex_set}

\subsection{Summary Statistics}

Our objective is to assess how effectively different summaries capture parameter-relevant information about underlying physical parameters by estimating their MI. This includes two aspects --- sufficiency (the \emph{total} information that summary statistics contain about the parameters), and complementarity (the \emph{additional} information that a summary provides beyond another baseline summary). 

For each application case, we compute the chosen summary statistics from the field data, fit a generative model to approximate the conditional density, and then use the fitted model to estimate MI and conditional MI between summaries and parameters. 

We consider two cases of application. The first case is a CMB-like GRFs. In principle, the PS contains all information in GRFs due to the Wick Theorem, thereby serving as a sanity check for MI as the metrics of information. The second application of MI is the 21~cm intensity mapping from the epoch of reionization (EoR), a non-Gaussian field for which there have been tremendous efforts to propose novel summary statistics in attempt to extract more information from the non-Gaussian 21~cm map. Here, we will use the MI to answer which statistics are more informative and investigate the complementarity between different summaries.  

Among many summaries, we select three representative statistics as examples in this paper, namely, the PS, the BS, and the wavelet ST. 
The PS is sufficient for GRFs and remains the most widely used statistic in cosmological analysis. 
The BS is the higher-order statistics in extension to the PS, which can capture the non-Gaussian features and is therefore widely employed in cosmological large-scale structure. Here, we use the reduced BS $Q(k_1,k_2,k_3)$ to separate out the Gaussian information as much as possible, but without loss of clarity, we still refer to reduced BS as BS throughout this paper. 
The wavelet ST is a more recent statistic inspired by signal processing and designed to capture hierarchical, multiscale non-Gaussian features. We recap their definitions in Appendix~\ref{app:summaries}. 

\subsection{Application I: CMB-like Gaussian Field}
\label{sec:CMB-ex}

\begin{figure}
    \centering
    \includegraphics[width=\linewidth]{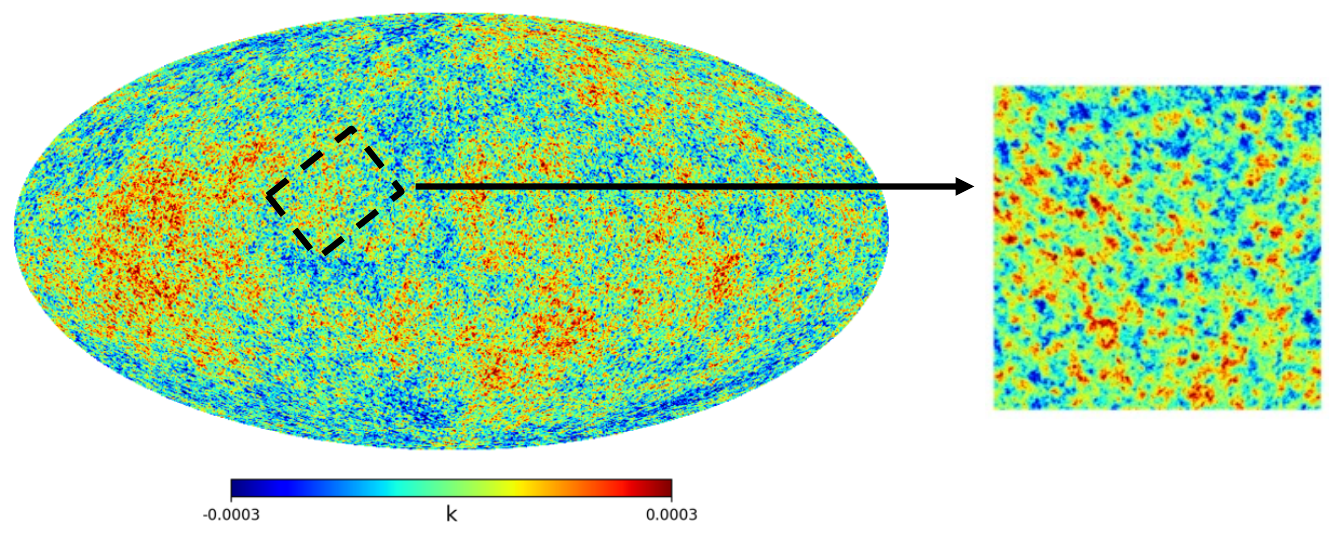}
    \caption{Left: a CMB-like map generated from a given PS. 
Right: a fixed square patch extracted from the map, used for computing summary statistics.}
    \label{fig:cmb_data}
\end{figure}

The information content of a GRF can, in principle, be fully characterized by its PS, 
making it an ideal test case for validating our framework. To construct an inference problem on GRFs, we adopt standard 
tools developed for CMB analysis. Specifically, we generate CMB power spectra with varying cosmological parameters using 
{\tt CAMB}\footnote{\url{https://github.com/cmbant/CAMB}} \citep{camb}, and produce corresponding CMB-like maps at 
${\tt nside}=1024$ with {\tt Healpy}\footnote{\url{https://github.com/healpy/healpy}} \citep{healpy1,healpy2}. 
By construction, these maps are pure GRFs. 
We extract a fixed square patch from each map, as illustrated in Figure~\ref{fig:cmb_data}, to calculate the statistics because some of these statistics are not defined directly on the sphere.

We vary two cosmological parameters, the dark matter density $\Omega_c$ and the Hubble parameter $h$. 
A $100 \times 100$ grid of parameter combinations is generated, with $\Omega_c \in [0.21, 0.30]$ and 
$h \in [0.64, 0.76]$, while all other cosmological parameters are fixed to the {\tt CAMB} defaults. 
For each parameter set, we simulate the corresponding map and compute a suite of summary statistics. 

To compute the PS and BS, we used the 2D polyspectra calculator implemented in the 
{\tt scattering}\footnote{\url{https://github.com/SihaoCheng/scattering_transform/blob/master/scattering/polyspectra_calculators}} 
package \citep{cheng2020new}. For the PS, we selected 15 logarithmically spaced $k$-bins within the range 
$[k_{\rm min}, k_{\rm max}]$, and used the same $k$-bins for the BS calculation. For the ST, we employed the Morlet wavelets implementation in 
{\tt Kymatio}\footnote{\url{https://www.kymat.io/}} \citep{2018arXiv181211214A}, with parameters $J=9$ and $L=4$, 
chosen to maximize scale coverage.

\begin{figure}
    \centering
    \includegraphics[width=\linewidth]{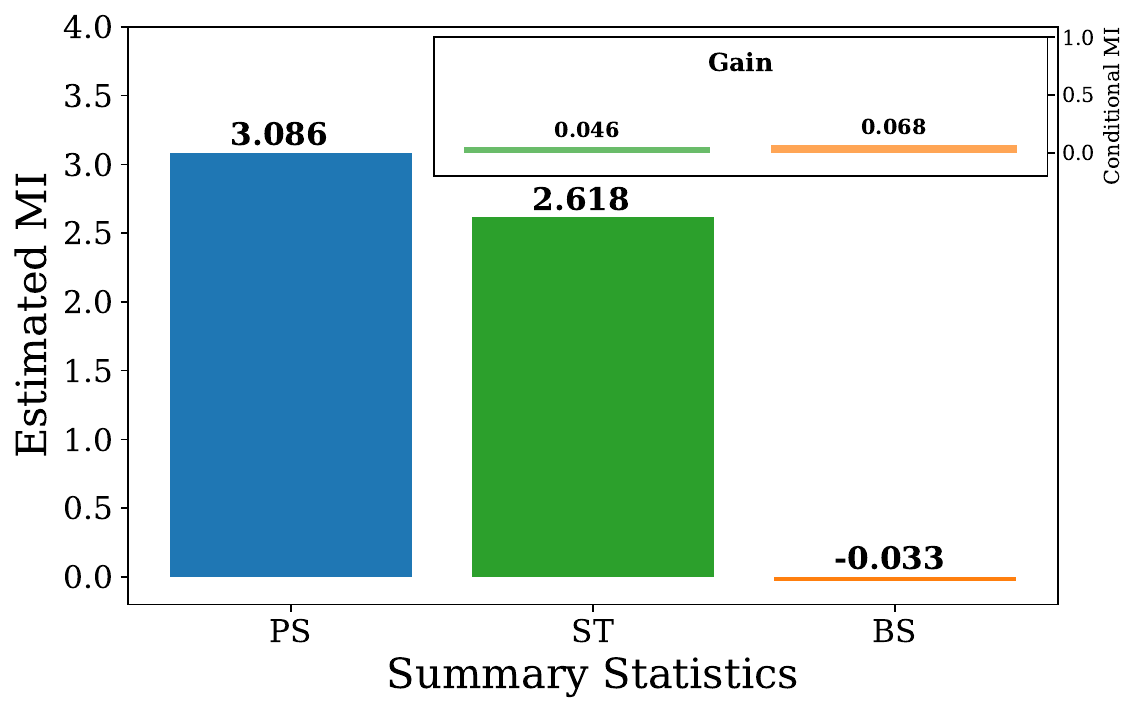}
    \caption{The MI of the PS (blue), ST (green) and BS (orange) for CMB-like GRFs. In the inset, the conditional MI of the ST (BS) given PS is shown, respectively, which measures the complementary information that the ST (BS) contributes beyond the PS.}
    \label{fig:mi_cmb}
\end{figure}

Figure~\ref{fig:mi_cmb} shows the information content of different summary statistics for CMB-like GRFs. The MI of the BS (-0.033), and the conditional MI of the BS (0.068) and ST (0.046) given the PS are negligible\footnote{The negative MI value of the BS indicates small numerical artifact in its estimation.}. This means that: (1) the BS contains almost no information from CMB-like GRFs; (2) while the MI of ST (2.618) is close to that of the PS (3.086), the ST contributes almost no additional information beyond the PS. In other words, the ST also captures the Gaussian information from the CMB-like GRFs. This test confirms that the MI and conditional MI correctly measure the information content in GRFs, which is consistent with the theoretical expectation that the PS contains all information in GRFs.

\subsection{Application II: non-Gaussian 21~cm Maps}
\label{sec:21cm_ex}

The 21~cm signal is non-Gaussian due to
reionization patchiness, an interesting application case for studying information sufficiency and complementarity. 


The 21~cm brightness temperature from the EoR relative to the CMB in position ${\bf x}$ is given by \citep{Furlanetto2006}:
\begin{equation}
T_{21}(\textbf{x},z)=\tilde{T}_{21}(z)\,x_{\rm HI}(\textbf{x})\,[1+\delta(\textbf{x})]\,
\left(1-\frac{T_{\rm CMB}}{T_S}\right)\,,
\end{equation}
where $\tilde{T}_{21}(z) = 27\sqrt{[(1+z)/10](0.15/\Omega_{\rm m} h^2)}(\Omega_{\rm b} h^2/0.023)$ mK, $x_{\rm HI}({\bf x})$ is the neutral hydrogen fraction, and $\delta({\bf x})$ is the matter overdensity. 
To simulate the cosmic 21~cm signal from the EoR, we generate coeval boxes using the publicly available {\tt 21cmFAST}\footnote{\url{https://github.com/andreimesinger/21cmFAST}} code \citep{Mesinger2007,Mesinger2011}, a semi-numerical simulation code for simulating the EoR 21~cm signal. Each simulation is run in a cubic volume of 128 comoving Mpc on each side, with $64^3$ grid cells. In principle, tomographic, lightcone 21~cm field that covers the entire EoR will be needed to measure all the information, but for the simple purpose of the demonstration of MI, we only simulate coeval boxes at redshift $z=12$. Also, we assume the spin temperature $T_S \gg T_\text{CMB}$, a valid approximation after the onset of the EoR. 

\begin{figure}
    \centering
    \includegraphics[width=\linewidth]{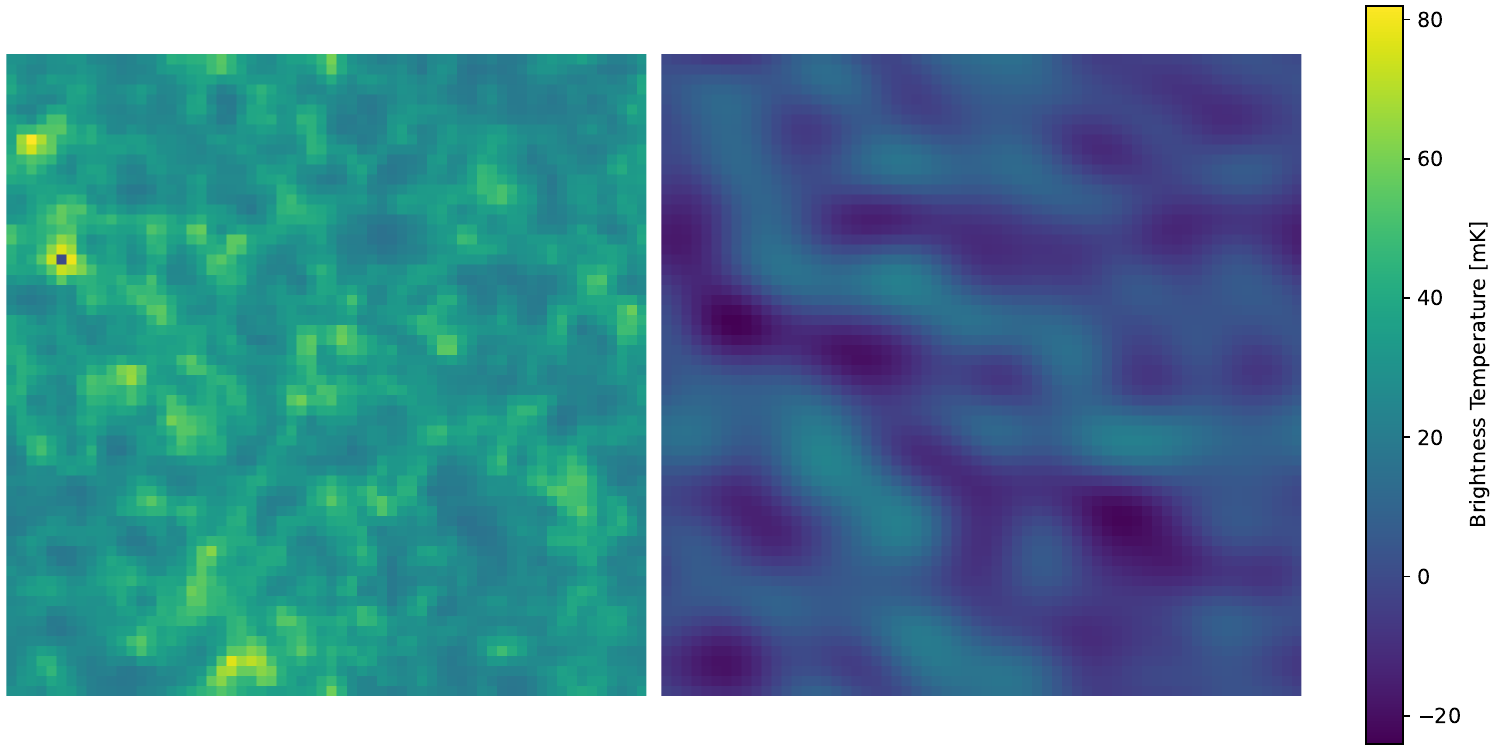}
    \caption{Illustration of the 21~cm maps. Left: cosmic 21~cm signal from the EoR. Right: mock observation (cosmic 21~cm signal with telescope noise) with SKA 1,000h observation.}
    \label{fig:21cm_data}
\end{figure}

We vary two reionization parameters, 
the ionizing efficiency $\zeta$ in the range $10 \le \zeta \le 250$, 
and the minimum virial temperature of halos hosting ionizing sources $T_{\mathrm{vir}}$ in the range $4 \le \log_{10}(T_{\mathrm{vir}}/\mathrm{K}) \le 6$. Cosmological parameters are fixed to the Planck results \citep{ade2016planck}, 
$(\Omega_{\Lambda}, \Omega_{m}, \Omega_{b}, n_s, \sigma_8, h) = (0.692, 0.308, 0.0484, 0.968, 0.815, 0.678)$.  


In addition to simulating the cosmic 21~cm signal from the EoR, we also consider the observational effects with the assumption of the Square Kilometre Array (SKA) configuration. We employ the {\tt Tools21cm}\footnote{\url{https://github.com/sambit-giri/tools21cm}} code \citep{tools21cm} to simulate instrumental noise for the SKA1-Low configuration. The $uv$ coverage is computed at the observed frequency corresponding to $z=12$, and used to model the telescope response and suppress thermal noise in $uv$ space. The system temperature is modeled as 
$T_{\rm sys}=60\,\left(\nu/300~{\rm MHz}\right)^{-2.55} + 100~{\rm K}$. Also, we assume a total of 1,000h observation, with 6 hours per day and 10~s integration time. The resulting noise cube is added to the signal cube to generate the mock observation. We illustrate an example of the cosmic 21~cm map and mock observation, respectively, in Figure~\ref{fig:21cm_data}.

\begin{figure}
    \centering
    \includegraphics[width=\linewidth]{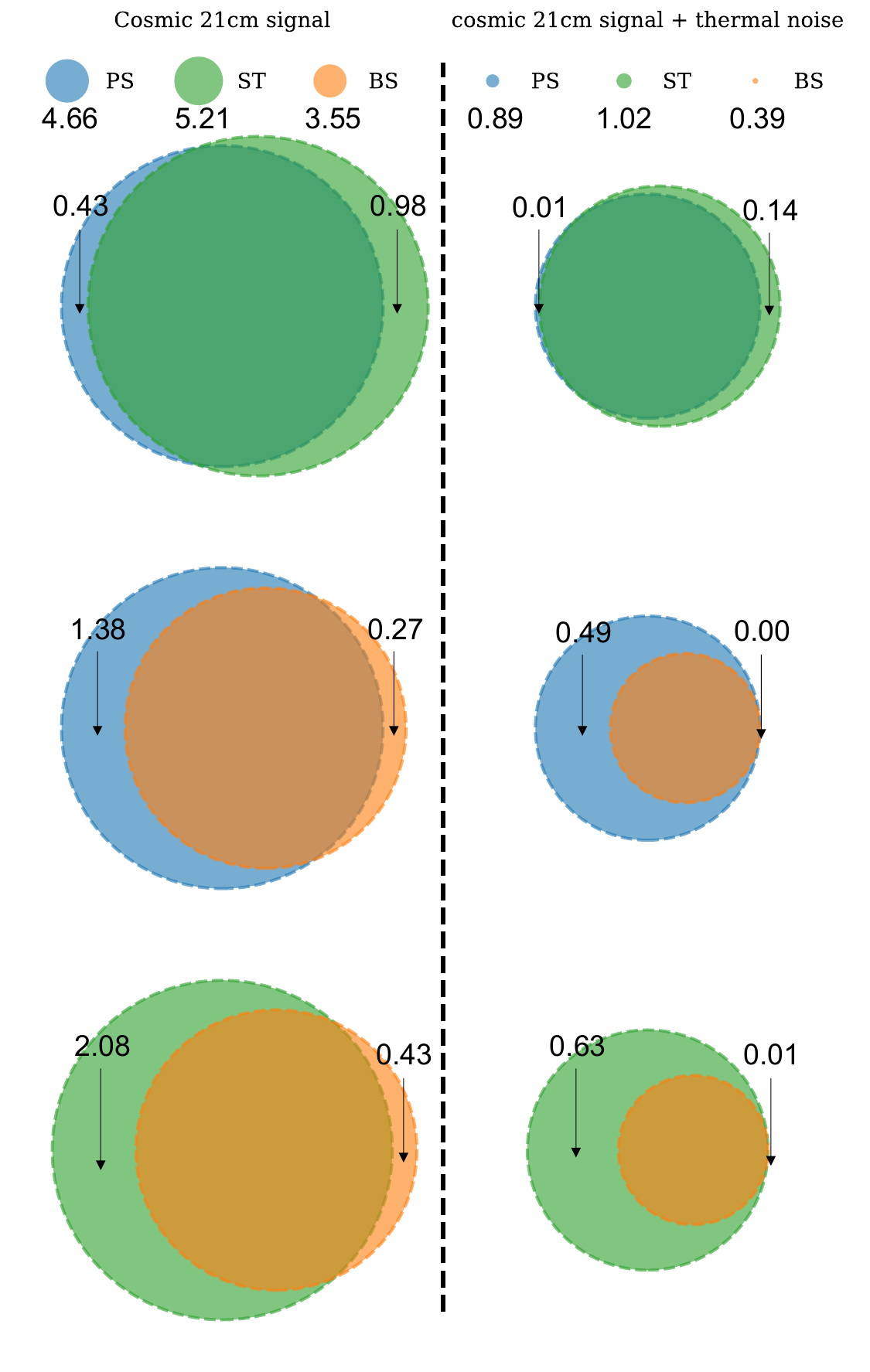}
    \caption{Venn diagram for the MI of summary statistics for the 21~cm maps from the EoR. We consider three summaries, PS (blue), ST (green) and BS (orange), and visualize their respective MI value (as marked) by the scaled circle size in the legend. We consider two cases, (left) the 21~cm map with only cosmic EoR signal, and (right) mock observations with 21~cm EoR signal and telescope noise with SKA 1,000h observation. For each case, we visualize the conditional MI (as marked with arrows) by the non-overlapped regions between two circles in Venn diagram.}
    \label{fig:mi_21cm}
\end{figure}



To calculate PS and BS, we employ the {\tt Tools21cm} code. Both PS and BS use the same 30 linearly uniform $k$-bins within $[k_{\rm min}, k_{\rm max}]$. For the ST, we use the {\tt Kymatio} code 
with solid harmonic wavelets. The parameters are $J=5$ scales, $L=5$ angles, coefficients up to second order, and modulus powers $q = [0.5,1,2]$. The coefficients are averaged over different orientations, following \citet{xs-st-delfi}.



In Figure~\ref{fig:mi_21cm}, we visualize the MI of three summary statistics for the 21~cm maps from the EoR and, in particular, visualize the conditional MI, i.e.\ complementary information of one summary beyond another, using the Venn diagram. 
In the case with only cosmic 21~cm signal (i.e.\ no thermal noise), while the PS contains substantial information (i.e.\ $I(\theta; {\rm PS}) = 4.66$), the ST captures the most (i.e.\ $I(\theta; {\rm ST}) = 5.21$). Here, $\theta = \lbrace \zeta, \log_{10}(T_{\mathrm{vir}}/\mathrm{K}) \rbrace$ is the set of reionization parameters. The ST contributes additional non-Gaussian information beyond the PS (i.e.\ $I(\theta; {\rm ST} \mid {\rm PS}) = 0.98$). In comparison, the BS extracts less information than ST and PS (i.e.\ $I(\theta; {\rm BS}) = 3.55$), and contributes less additional information beyond the PS or ST (i.e.\ $I(\theta; {\rm BS} \mid {\rm PS}) = 0.27$ and $I(\theta; {\rm BS} \mid {\rm ST}) = 0.43$). 

Note that each pair of PS, ST and BS has nonzero conditional MI beyond each other for the cosmic 21~cm field. This implies that these three summaries extract different, albeit not completely different, aspects of information which are indeed complementary to each other.

For the case of mock observation that is applied with thermal noise assuming SKA 1,000h observation, the MI of these summaries for the mock observation is significantly smaller (i.e.\ $I(\theta; {\rm PS}) = 0.89$, $I(\theta; {\rm ST}) = 1.02$ and $I(\theta; {\rm BS}) = 0.39$) than that for the case of cosmic signal. Basically, thermal noise is independent of reionization parameters, suppressing the inference effectiveness of the total 21~cm signal.

Note that the conditional MI, $I(\theta; {\rm PS} \mid {\rm ST})$, $I(\theta; {\rm BS} \mid {\rm PS})$ and $I(\theta; {\rm BS} \mid {\rm ST})$, are all nearly zero in this case. At face value, this means that the information of the BS is a subset of the PS, and the information of the PS is also a subset of the ST, when thermal noise at the SKA sensitivity level is applied. 
This interesting result serves as a proof-of-concept for MI as a metric of information content under our simplified simulations and mock observations, but it should not be overly interpreted. For example, the MI of BS might be underestimated herein because the calculation of BS only takes into account the configuration of equilateral triangles. More realistic assumptions on instrumental effects and systematics than in this work might also find a different complementarity between these summaries (see, e.g.\ \citealt{Watkinson_bs_eor_infer}).

In summary, for all cases, the ST successfully extracts the most information (see also, \citealt{xs-st-delfi}), the PS ranks second, and the BS provides the least information.
For the cosmic 21~cm map which is non-Gaussian, ST also provides significantly complementary information beyond the PS, whereas the BS adds a little beyond the PS. For the mock 21~cm observation with thermal noise, the information extracted by the ST contains that by the PS which also contains that by the BS. 




These results demonstrate that the MI framework can successfully provide a quantitative assessment of both the total and complementary information content of different summary statistics. In addition, the application to the 21~cm map highlights the advantage of ST in capturing non-Gaussian features from the 21~cm data.

\section{Conclusions}
\label{sec:conclusion}
In this paper, we demonstrate that MI provides a powerful and principled metric for evaluating the effectiveness of summary statistics in cosmological inference. Using GRFs as a validation case, we confirm that MI correctly identifies the optimal summaries. We then applied the approach to 21~cm simulations, showing how MI can quantify both the absolute information content of different summaries and their complementary contributions, enabling the systematic selection of summaries that enhance a base set.

Beyond evaluation, MI offers a natural target for learning informative summaries. By incorporating MI into loss functions, neural networks can be trained to produce summaries that maximize relevant information for inference tasks, offering a path toward automated and optimized summary design.

Current MI estimators, however, face challenges in very high-dimensional settings, such as field-level data. Advancing estimation techniques in this regime will allow us to fully quantify the maximal extractable information from complex cosmological fields, bridging the gap between theory and practical data analysis. Overall, MI provides both a theoretical foundation and a practical framework for designing and assessing summary statistics in modern cosmology.

\begin{acknowledgments}
This work is supported by National SKA Program of China (grant No.~2020SKA0110401). X.Z. acknowledges the support through a grant from the Schmidt Sciences. We acknowledge the Tsinghua Astrophysics High-Performance Computing platform at Tsinghua University for providing computational and data storage resources that have contributed to the research results reported within this paper. 
\end{acknowledgments}

\software{21cmFAST \citep{Mesinger2007,Mesinger2011}, NumPy \citep{harris2020array}, Matplotlib \citep{Hunter:2007}, SciPy \citep{2020SciPy-NMeth}, scikit-learn \citep{Scikit-learn}, Python3 \citep{py3}, Kymatio \citep{2018arXiv181211214A}, Tools21cm \citep{tools21cm},healpy \citep{healpy1,healpy2},camb \citep{camb},ltu-ili\citep{ltu-ili}}, Pylians \citep{Pylians}


\appendix

\section{MI and Bayesian Sufficiency}
\label{app:proof}
Let $s=f(x)$ be a deterministic statistic. Then $(\theta \to x \to s)$ forms a Markov chain. By the chain rule for MI,
\begin{equation}
I(\theta; x,s) = I(\theta; s) + I(\theta; x \mid s).
\end{equation}
Since $s$ is a function of $x$, we also have $I(\theta; x,s)=I(\theta;x)$. Hence, 
\begin{equation}
I(\theta; x) = I(\theta; s) + I(\theta; x \mid s).
\label{eq:chain}
\end{equation}
Thus $I(\theta; x)=I(\theta; s)$ if and only if $I(\theta; x\mid s)=0$. Now, $I(\theta; x\mid s)=0$ holds if and only if $\theta \perp x \mid s$, i.e.
\begin{equation}
p(\theta \mid x, s) = p(\theta \mid s) 
\end{equation}
Because $s=f(x)$, conditioning on $(x,s)$ is the same as conditioning on $x$, so $p(\theta\mid x,s)=p(\theta\mid x)$. Therefore
\begin{equation}
I(\theta; x\mid s)=0
\;\;\Longleftrightarrow\;\;
p(\theta\mid x)=p(\theta\mid s(x)) 
\label{eq:suff}
\end{equation}
Combining \eqref{eq:chain} and \eqref{eq:suff} yields
\[
p(\theta\mid x)=p(\theta\mid s(x)) 
\quad\Longleftrightarrow\quad
I(\theta; x)=I(\theta; s(x)).
\]
\hfill$\Box$

\section{Relationships between MI, FI, and Estimation MSE}
\label{app:FI_MI_ineq}

For a likelihood $p(x|\theta)$, the FI matrix is defined as
\begin{equation}
\label{eq:fim}
\mathcal{I}_F(\theta) \equiv \mathbb{E}_{p(x|\theta)}\!\left[ \nabla_\theta \log p(x|\theta) \, \nabla_\theta \log p(x|\theta)^\top \right].
\end{equation}
According to the Cramér–Rao bound, no unbiased estimator can achieve a smaller variance than the inverse of the FI matrix. This makes FI a natural tool to quantify how much information an observable carries about the parameters.

For a likelihood model $p(x|\theta)$ and a smooth prior $p(\theta)$, in addition to the FI matrix defined in Equation~(\ref{eq:fim}), we define the \emph{prior information matrix} as
\begin{equation}
\mathcal{J} \equiv \mathbb{E}_{p(\theta)}\!\left[ \nabla_\theta \log p(\theta) \, \nabla_\theta \log p(\theta)^\top \right],
\end{equation}
which quantifies the informativeness of the prior. 

Efroimovich's inequality \citep{info_ineq} states that
\begin{equation}
\label{eq:Ef_ineq}
\frac{1}{2\pi e} \, e^{\frac{2}{d} h(\theta|x)} \ge \frac{1}{\det\Big( \mathcal{J} + \mathbb{E}_{p(\theta)}[\mathcal{I}_F(\theta)] \Big)^{1/d}},
\end{equation}
or
\begin{equation}
\label{eq:Ef_ineq2}
-\frac{2}{d} \,h(\theta|x) \le \log\left[ (2\pi e)^{-1} {\det\Big( \mathcal{J} + \mathbb{E}_{p(\theta)}[\mathcal{I}_F(\theta)] \Big)^{1/d}} \right]\,,
\end{equation}
where $d$ is the dimension of $\theta$. Here, $h(\theta|x) \equiv -\mathbb{E}_{p(\theta,x)}[\log p(\theta|x)]$ is the conditional entropy, and we have $I(\theta ; x) = - h(\theta|x) + h(\theta)$, where $h(\theta)$ is the differential entropy of $\theta$. Therefore, ignoring $h(\theta)$ (since it is independent of the field $x$), the Efroimovich’s inequality implies that the FI integrated over the prior $p(\theta)$ provides an upper bound on the MI. 

On the other hand, assume we have an estimator $\hat{\theta}(x)$ of the parameter $\theta$. Using the data processing inequality and the fact that entropy is maximized by a Gaussian, we have
\begin{align}
\frac{2}{d} \, h(\theta|x) 
&\le \frac{2}{d} \, h(\hat{\theta}-\theta) \notag\\
&\le \log\Big( (2\pi e) \det(\Sigma)^{1/d} \Big) \notag\\
&\le \log\Bigg[ (2\pi e) \frac{1}{d} \, \mathbb{E}_{p(\theta,x)} \|\hat{\theta}-\theta\|^2 \Bigg]\,,
\end{align}
or
\begin{equation}
-\frac{2}{d} \, h(\theta|x) \ge 
-\log\Bigg[ (2\pi e) \frac{1}{d} \, \mathbb{E}_{p(\theta,x)} \|\hat{\theta}-\theta\|^2 \Bigg]\,.
\label{eq:vantree2}
\end{equation}
This means that the Bayesian estimation error provides a lower bound on the MI. 
These inequalities (Equations~\ref{eq:Ef_ineq2} and \ref{eq:vantree2}) reveal interesting relationships between different information metrics: MI is upper bounded by FI, and lower bounded by the mean squared error (MSE) of the estimation. 

Combining Equations~(\ref{eq:Ef_ineq2}) and (\ref{eq:vantree2}) yields
\begin{equation}
\mathbb{E}_{p(\theta,x)}\|\hat{\theta}-\theta\|^2\ge \frac{d}{\det\Big( \mathcal{J} + \mathbb{E}_{p(\theta)}[\mathcal{I}_F(\theta)] \Big)^{1/d}}\,.
\end{equation}
This is known as \emph{van Trees inequality} \citep{VanTrees_ineq,non_parametric_est}, or the Bayesian Cramér–Rao bound \citep{BCR_bound}.  

\section{Summary Statistics}
\label{app:summaries}

We summarize the definitions and main configuration parameters of summaries used in this paper.

\subsection{PS}
The PS $P(\boldsymbol{k})$ is the most widely used statistic in cosmology. For a 3D field, it is defined as
\begin{equation}
    \left\langle\delta(\boldsymbol{k}) \delta\left(\boldsymbol{k}^{\prime}\right)\right\rangle 
    = (2 \pi)^3 \delta^D\left(\boldsymbol{k}+\boldsymbol{k}^{\prime}\right) P(\boldsymbol{k}),
\end{equation}
where $\delta(\boldsymbol{k})$ is the Fourier transform of the field, $\delta^D$ is the Dirac delta function, and $\langle\cdot\rangle$ denotes an ensemble average.  

In practice, ensemble averaging is replaced by binning over spherical shells in Fourier space, assuming statistical isotropy:
\begin{equation}
    P(k) \approx 
    \frac{\sum_{\boldsymbol{k} \in k} \left|\delta(\boldsymbol{k})\right|^2}{N_k V},
\end{equation}
where $V$ is the survey volume and $N_k$ is the number of modes in the spherical $k$-shell $[k-\Delta k/2, k+\Delta k/2]$. This procedure yields the spherically averaged power spectrum. While spherical harmonic methods are often used for CMB and other sky-projected fields, in this paper we restrict our PS calculations to 2D and 3D flat fields.  

The key parameters for the PS calculation are the binning scheme and the accessible $k$-range. For a $d$-dimensional field with box size $L$ and $N$ grid cells per dimension, the minimum and maximum Fourier modes are
\begin{equation}
    k_{\min} = \frac{2\pi}{L}, \qquad k_{\max} = \frac{\pi N \sqrt{d}}{L}.
\end{equation}
The bins are typically spaced either linearly or logarithmically within this range. 

\subsection{BS}
Non-Gaussian cosmological fields motivate the use of higher-order correlation functions. The BS is the Fourier transform of the three-point correlation function. It is defined as
\begin{align}
    \left\langle\delta(\boldsymbol{k}_1)\, \delta(\boldsymbol{k}_2)\, \delta(\boldsymbol{k}_3)\right\rangle 
    = & (2 \pi)^3 \delta^D\left(\boldsymbol{k}_1+\boldsymbol{k}_2+\boldsymbol{k}_3\right) \nonumber \\ 
    & \times B(\boldsymbol{k}_1,\boldsymbol{k}_2,\boldsymbol{k}_3)\,.
\end{align}
Similar to PS, the bispectra are spherically averaged over triangle orientations thanks to isotropy, so that the BS depends only on triangle shape and size. The reduced BS is also defined to separate information from the PS, 
\begin{equation}
Q(k_1, k_2, k_3) \equiv 
\frac{B(k_1, k_2, k_3)}
{P(k_1)P(k_2) + P(k_2)P(k_3) + P(k_3)P(k_1)} .
\end{equation}

In general, the BS requires sampling over all closed triangles of wavevectors, which is computationally demanding. For tractability, in this paper we restrict ourselves to the configuration of equilateral triangles, parameterized by a single $k$ value. This choice keeps the parameter dependence consistent with that of the PS, with the key parameter choice again being the $k$-binning.

\subsection{ST}
Introduced by \citet{mallat11}, the ST provides a hierarchical representation of fields by iteratively combining wavelet convolutions, nonlinearities, and spatial averaging. For a field $x$, the order-$m$ scattering coefficient at scales $(\lambda_1,\ldots,\lambda_m)$ is
\begin{equation}
    Sx[\lambda_1,\ldots,\lambda_m] 
    = |\psi_{\lambda_m}\star \cdots |\psi_{\lambda_1}\star x|\cdots|,
\end{equation}
where $\star$ denotes convolution and $\{\psi_\lambda\}$ is a family of wavelets. For 3D fields, an additional dimensionality reduction step is used, in which the modulus outputs are raised to the power $q$ and spatially averaged into scalar coefficients.  

The main configuration parameters of the ST include (1) the choice of wavelet family (e.g., Morlet, solid harmonic); (2) the index for the scale $J$ and the orientation $L$; (3) the maximum order $m$ (typically 1 or 2); (4) the modulus exponent $q$, controlling sensitivity to large vs small fluctuations. 
These hyperparameters jointly determine the richness and dimensionality of the resulting summary.

The ST captures multiscale, multiorientation features while enforcing translation and rotation invariance. It has been demonstrated to recover non-Gaussian information and has applied to parameter inference in 3D reionization fields \citep{cheng2020new, xs-st-delfi}.

\bibliography{ref}{}
\bibliographystyle{aasjournalv7}



\end{document}